\documentclass[aps,prc,showpacs,twocolumn,nofootinbib,groupedaddress,preprintnumbers]{revtex4-1}

\usepackage{color}
\usepackage{calligra}
\usepackage{graphicx}
\usepackage{dcolumn}
\usepackage{bm}
\usepackage{amsmath,amssymb}
\usepackage{mathrsfs}
\usepackage{subfigure}

\newcommand{\ket}[1]{| \, #1 \, \rangle}

\begin{document}

\preprint{YITP-18-18}


\title{Theoretical study of the $\Lambda(1405)$ resonance in $\Xi_b^{0}\to D^{0}(\pi\Sigma) $ decay}


\author{Kenta~Miyahara}
\email[]{miyahara.kenta.62r@st.kyoto-u.ac.jp}
\affiliation{Department of Physics, Graduate School of Science, Kyoto University, Kyoto 606-8502, Japan}
\author{Tetsuo~Hyodo}
\email[]{hyodo@yukawa.kyoto-u.ac.jp}
\affiliation{Yukawa Institute for Theoretical Physics, Kyoto University, Kyoto 606-8502, Japan}


\date{\today}

\begin{abstract}       
We study the mechanism of the weak decay process of $\Xi_b^{0} $ into $D^{0}$ with a meson-baryon pair. It is shown that the dominant component of the prompt weak decay produces the meson-baryon pair with the spectator strange quark being transferred to the baryon, so that the $\bar{K}N$ channel is absent. Subsequent final state interaction then reflects the $\pi\Sigma$ originated $\Lambda(1405)$ formation amplitude, which has been difficult to access in previous experimental studies. We predict the line shapes of the $\pi\Sigma$ invariant mass distribution using a realistic chiral meson-baryon amplitude for the final state interaction. It is shown that the interference between the direct and the rescattering contributions can strongly distort the peak structure of the $\Lambda(1405)$ in the mass distribution. This indicates the necessity of a detailed investigation of the reaction mechanism in order to extract the $\Lambda(1405)$ property in the $\pi\Sigma$ mass distribution.

\end{abstract}




\maketitle

\section{Introduction}   \label{sec:intro}  

The $\Lambda(1405)$ resonance has attracted the attention of researchers since its discovery~\cite{Dalitz:1959dn,Dalitz:1960du,Alston:1961zz}. In addition to the traditional problem in the constituent quark model~\cite{Isgur:1978xj}, several other issues have arisen in the study of the $\Lambda(1405)$~\cite{Hyodo:2011ur,Kamiya:2016jqc}. For instance, because it is located close to the $\bar{K}N$ threshold, the properties of the $\Lambda(1405)$ have a substantial influence on the $\bar{K}N$ interaction~\cite{Hyodo:2007jq,Miyahara:2015bya}, which is the fundamental building block to construct $\bar{K}$ bound states in nuclei~\cite{Akaishi:2002bg,Gal:2016boi}. Furthermore, through the development of chiral unitary approaches~\cite{Kaiser:1995eg,Oset:1998it,Oller:2000fj}, it is shown that there exist two resonance poles of the scattering amplitude in the $\Lambda(1405)$ energy region~\cite{Oller:2000fj,Jido:2003cb}. The existence of the two poles was first unveiled in Ref.~\cite{Oller:2000fj}, and the origin of two poles in the flavor SU(3) symmetric limit and their implication on the $\pi\Sigma$ invariant mass spectrum were discussed in Ref.~\cite{Jido:2003cb}. Because a pole of the amplitude represents an eigenstate of the system, this indicates that the $\Lambda(1405)$ is not a single resonance but a superposition of two resonances. The existence of two poles is confirmed by the recent analyses~\cite{Ikeda:2011pi,Ikeda:2012au,Guo:2012vv,Mai:2014xna} including the precise constraint from the kaonic hydrogen measurement~\cite{Bazzi:2011zj,Bazzi:2012eq}, and is now tabulated in the listings by the Particle Data Group~\cite{Olive:2016xmw}.

From the viewpoint of the experimental analysis, the meson-baryon two-body scattering amplitude in the $\Lambda(1405)$ region is not directly obtained. Because the $\Lambda(1405)$ lies below the $\bar{K}N$ threshold, the only open channel is $\pi\Sigma$, for which the direct scattering experiment is not possible. It is therefore needed to perform production experiments, whose final state consists of several particles in addition to the $\Lambda(1405)$. To extract the information of the meson-baryon scattering amplitude in such reactions, theoretical models have to be employed to describe the reaction mechanism. There are several theoretical studies on the mechanism of various production reactions, such as $\gamma p\to K^{+}\Lambda(1405)$~\cite{Nacher:1998mi,Nakamura:2013boa,Roca:2013av,Roca:2013cca,Wang:2016dtb}, $\gamma p\to K^{*+}\Lambda(1405)$~\cite{Hyodo:2004vt}, $\bar{\nu}_{\mu}p\to \mu^{+}\Lambda(1405)$~\cite{Ren:2015bsa}, $K^{-} p\to \gamma\Lambda(1405)$~\cite{Nacher:1999ni}, $\pi^{-}p\to K^{0}\Lambda(1405)$~\cite{Hyodo:2003jw,Bayar:2017svj}, $K^{-}p\to \pi^{0}\Lambda(1405)$~\cite{Magas:2005vu}, $pp\to K^{+}\Lambda(1405)$~\cite{Geng:2007vm,Bayar:2017svj}, and $K^{-}d\to n\Lambda(1405)$~\cite{Jido:2009jf,Jido:2010rx,Miyagawa:2012xz,Revai:2012fx,Jido:2012cy,Ohnishi:2015iaq,Kamano:2016djv,Miyagawa:2018xge}. In general, the invariant mass distribution of the $\pi\Sigma$ state depends on a particular superposition of the $\bar{K}N\to \pi\Sigma$ and $\pi\Sigma\to \pi\Sigma$ amplitudes, leading to the different peak structure of the $\Lambda(1405)$.

Recently, yet another class of experimental method has been proposed to study hadron resonances through the final state interactions of the weak decay of heavy hadrons~\cite{Oset:2016lyh}. In fact, the study of the $\Lambda(1405)$ has been performed in the $\Lambda_{b}\to J/\psi \Lambda(1405)$ decay~\cite{Roca:2015tea}, the $\Lambda_{c}\to \pi^{+}\Lambda(1405)$ decay~\cite{Miyahara:2015cja}, the $\Lambda_{c}\to \nu l^{+}\Lambda(1405)$ decay~\cite{Ikeno:2015xea}, and the $\Lambda_{b}\to \eta_{c}\Lambda(1405)$ decay~\cite{Xie:2017gwc}. In particular, it is shown in Ref.~\cite{Miyahara:2015cja} that in the dominant quark diagram of the $\Lambda_{c}\to \pi^{+}MB$ process, the strange quark from the charm quark decay is selectively transferred to the meson state $M$, so that the $MB$ pair does not contain the $\pi\Sigma$ channel. This is ideal to study the $\bar{K}N\to \pi\Sigma$ amplitude for the $\Lambda(1405)$ formation.
It would therefore be helpful to find a process where the $\pi\Sigma\to \pi\Sigma$ amplitude dominates the final state interaction.

In this paper, we study the $\Xi_b^{0}\to D^{0}MB$ decay where $MB$ stands for the meson-baryon pair having the same quantum number with the $\Lambda(1405)$. The key idea is that, in the dominant decay process, the strange quark in the initial $\Xi_{b}^{0}$ state should be transferred to the baryon in the final state as a spectator. As a consequence, the weak decay does not produce the $\bar{K}N$ state where the baryon has no strangeness. In the next section, we discuss the dominant weak decay mechanism which realizes the above idea, along the same line as Refs.~\cite{Roca:2015tea,Miyahara:2015cja}. We then show the prediction of the invariant mass distribution of the $MB$ states in Sec.~\ref{results}, using the reliable final state interaction model~\cite{Ikeda:2011pi,Ikeda:2012au}. The last section is devoted to summary.

\section{Formulation}    \label{sec:formulation}  

Following the formulation in Refs.~\cite{Roca:2015tea,Miyahara:2015cja,Oset:2016lyh}, we study the invariant mass distribution of $\Xi_b^0\to D^0\Lambda(1405)\to D^0(\pi\Sigma)$. We consider the prompt weak decay process $\Xi_{b}^{0}\to D^{0}MB^{\prime}$ and the strong final state interaction $MB^{\prime}\to MB$ separately.

\begin{figure*}[tb]
\begin{center}
\subfigure[]{
\includegraphics[width=5.5cm,bb=0 0 400 303]{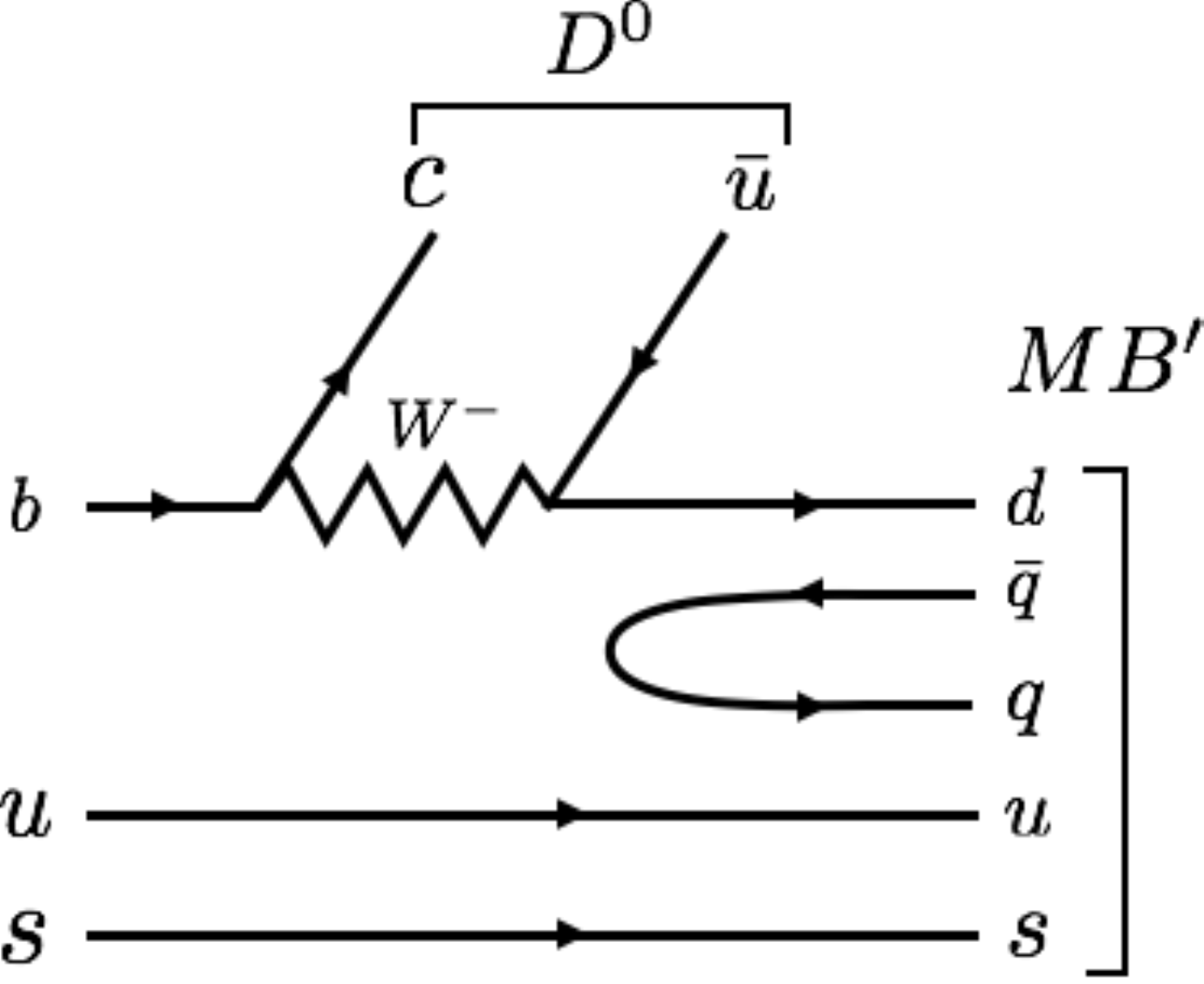}
}
\subfigure[]{
\includegraphics[width=5.5cm,bb=0 0 447 234]{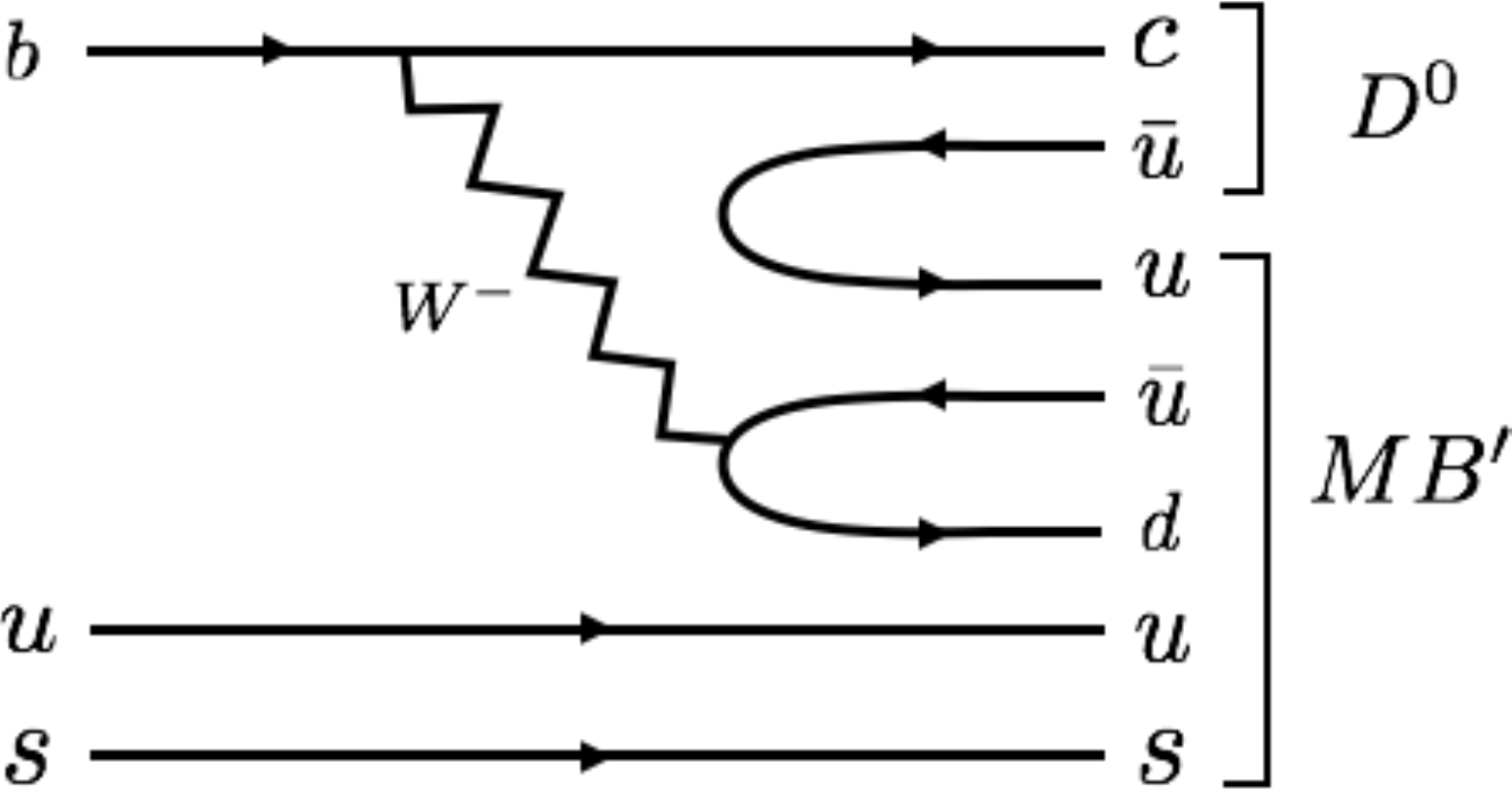}
}
\subfigure[]{
\includegraphics[width=5.5cm,bb=0 0 447 234]{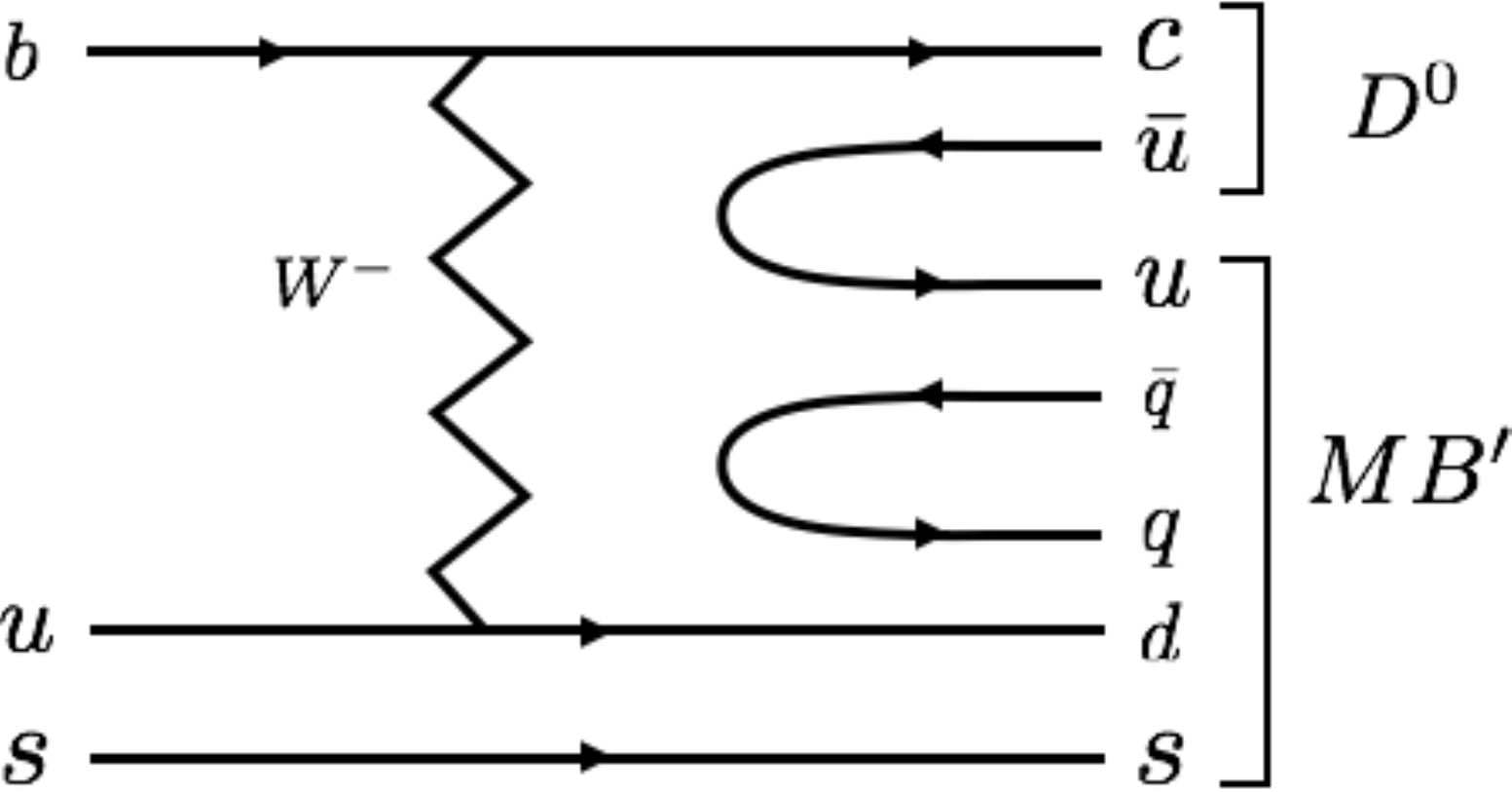}
}
\caption{Quark-line diagrams of the Cabibbo favored mechanisms for the $\Xi_{b}^{0}\to D^{0}MB^{\prime}$ process. The solid and wiggly lines denote the quark and $W$ boson, respectively.
}
\label{fig:weak}  
\end{center}
\end{figure*}

In the first step of the $\Xi_{b}^{0}\to D^{0}MB^{\prime}$ process, the Cabibbo favored transitions occur through the $b$ decay ($b\to c\bar{u}d$) or the $bu$ scattering ($bu\to cd$). Possible quark-line diagrams for $\Xi_{b}^{0}\to D^{0}MB^{\prime}$ are shown in Fig.~\ref{fig:weak}. To pin down the dominant diagram of this decay, we consider the kinematics of the process. Because we are interested in the final $MB^{\prime}$ pair in the $\Lambda(1405)$ energy region, the emitted $D^0$ meson should have a high momentum ($\sim$2~GeV in the rest frame of the initial $\Xi_b^0$) due to the heavy mass of $\Xi_{b}$ ($\sim$5.8~GeV). In the diagram in Fig.~\ref{fig:weak}(a), both the $c$ and $\bar{u}$ quarks in the $D^0$ are produced directly from the $b$ decay, which is favorable for the high-momentum $D^0$ emission. On the other hand, in Figs.~\ref{fig:weak}(b) and \ref{fig:weak}(c), the $\bar{u}$ quark in the $D^{0}$ is created from the vacuum, which is considered to be the soft mechanism with the momentum around a few hundred MeV. Figures~\ref{fig:weak}(b) and \ref{fig:weak}(c) are therefore suppressed for the mismatch of the kinematical condition to produce the high-momentum $D^0$. The diagram in Fig.~\ref{fig:weak}(c), containing the $bu$ scattering (called ``absorption diagrams'' in Ref.~\cite{Chau:1982da}), is further suppressed as discussed in Ref.~\cite{Xie:2014tma}, because of the additional $\bar{q}q$ creation and the two-body nature of the process. With these considerations, we conclude that Fig.~\ref{fig:weak}(a) is the dominant diagram in the $\Xi_b^0\to D^0MB^{\prime}$ process in the present kinematics and neglect the others in the following discussion. 

%
%

The intermediate state $|MB^{\prime}\rangle$ is a superposition of several meson-baryon channels with strangeness $S=-1$. To derive the relative weight of the channels, we express the process in Fig.~\ref{fig:weak}(a) in terms of the quark degrees of freedom. The initial $\Xi_b$ belongs to the flavor ${\bf \bar{3}}$ configuration~\cite{Roberts:2007ni}, 
\begin{align}
|\Xi_b^0\rangle = \frac{1}{\sqrt{2}}|b(su-us)\rangle.
\end{align}
After the $b$ decay, $b\to c\bar{u}d$, it leads to $\sim |c\bar{u}d(su-us)\rangle\sim |D^{0}d(su-us)\rangle$. Because the $D^0$ from the $c\bar{u}$ pair has the extremely high momentum and can be neglected in the following discussion, we concentrate only on the remaining part, 
\begin{align}
|MB^\prime\rangle \sim \frac{1}{\sqrt{2}}|d(su-us)\rangle.
\label{eq:uds}
\end{align}
The meson-baryon fractions can be read off by adding a $q\bar{q}$ pair in this wave function. As discussed in Refs.~\cite{Miyahara:2015cja,Miyahara:2016yyh}, the $q\bar{q}$ creation should be attached to the $d$ quark from the $b$ decay, because of the angular momentum matching and the strong $us$ diquark correlation in $\Xi_{b}$ (known as the good diquark~\cite{Jaffe:2004ph}). Thus, the meson is composed of the $d$ quark and the created antiquark, and the baryon of the $us$ diquark and the created quark. This is achieved by inserting the flavor singlet $q\bar{q}$ pair in Eq.~\eqref{eq:uds} as 
\begin{align}
|MB^\prime\rangle 
&=\frac{1}{\sqrt{2}}\sum_{i=1}^{3} |d\bar{q}_{i}q_i(su-us)\rangle \\
&=\frac{1}{\sqrt{2}}\sum_{i=1}^{3} |P_{2i}q_i(su-us)\rangle ,
\end{align}
with
\begin{align}
q&\equiv
\left( \begin{array}{c}
u \\ d \\ s
\end{array} \right), \quad
P\equiv q\bar{q}=
\left( \begin{array}{ccc}
u\bar{u}  &  u\bar{d}  &  u\bar{s}  \\
d\bar{u}  &  d\bar{d}  &  d\bar{s}  \\
s\bar{u}  &  s\bar{d}  &  s\bar{s}  
\end{array} \right).  \notag
\end{align}
To express this $|MB^{\prime}\rangle$ state in terms of the hadronic degrees of freedom, we represent the $P$ matrix by the pseudoscalar meson nonet
\begin{align}
P=\left( \begin{array}{ccc}
\frac{\pi^0}{\sqrt{2}}+\frac{\eta}{\sqrt{3}}+\frac{\eta^\prime}{\sqrt{6}}  &  \pi^+  &  K^+  \\
\pi^-  &  -\frac{\pi^0}{\sqrt{2}}+\frac{\eta}{\sqrt{3}}+\frac{\eta^\prime}{\sqrt{6}}  &  K^0  \\
K^-  &  \bar{K}^0  &  -\frac{\eta}{\sqrt{3}}+\frac{2\eta^\prime}{\sqrt{6}}
\end{array} \right).  \notag
\end{align}
The remaining three quarks should be represented by the ground state baryons. Using the phase convention of the quark representation of the baryon states in Ref.~\cite{Miyahara:2016yyh}, we obtain the following expression:
\begin{align}
|MB^\prime\rangle &=
-\frac{1}{2\sqrt{3}}|\pi^{0}\Lambda\rangle 
+\frac{1}{2}|\pi^0\Sigma^0\rangle 
+|\pi^-\Sigma^+\rangle 
 \notag \\
&\quad +\frac{1}{3\sqrt{2}}|\eta\Lambda\rangle 
-\frac{1}{\sqrt{6}}|\eta\Sigma^{0}\rangle
+|K^{0}\Xi^{0}\rangle ,  \label{eq:MB}
\end{align}
where we have neglected the $\eta^{\prime}\Lambda$ channel whose threshold  is much higher than the energy region of interest. Equation~\eqref{eq:MB} determines the relative weight of the meson-baryon channels in this weak decay. It should be noted that the $\ket{MB^{\prime}}$ state does not contain the $\bar{K}N$ state, because the $s$ quark in the initial $\Xi_{b}^{0}$ is transferred to the baryon. This is in sharp contrast to the $\Lambda_{c}\to \pi MB$ decay where the $\pi\Sigma$ state is absent~\cite{Miyahara:2015cja}.

Next, we consider the final state interaction of $MB^{\prime}\to MB$. A suitable framework is the chiral unitary approaches~\cite{Kaiser:1995eg,Oset:1998it,Oller:2000fj}, in which the $\Lambda(1405)$ is dynamically generated by the meson-baryon dynamics. In this study, we utilize the model in Refs.~\cite{Ikeda:2011pi,Ikeda:2012au} where the low-energy $K^{-}p$ scattering observables are systematically fitted with the accuracy of $\chi^{2}/{\rm d.o.f.}\sim 1$, by using the next-to-leading-order interaction kernel in chiral perturbation theory. With this final state interaction model, the transition amplitude of $\Xi_{b}\to D^{0}(MB)_{j}$ is given by
\begin{align}
\mathscr{M}_j =V_P &\left( h_j + \sum_i h_i G_i(M_{\rm inv})T_{ij}(M_{\rm inv}) \right),  \label{eq:amplitude} 
\end{align}
where $V_{P}$ represents the strength of the weak decay process shown in Fig.~\ref{fig:weak}(a), $T_{ij}$ is the coupled-channel meson-baryon scattering amplitude, $G_{i}$ is the two-body loop function, and $M_{\rm inv}$ is the invariant mass of the $MB$ state. The weight factor $h_{i}$
is determined by Eq.~\eqref{eq:MB} as
\begin{align}
h_{\pi^0\Lambda^0}
&=-\frac{1}{2\sqrt{3}} , \notag \\
h_{\pi^0\Sigma^0}
&=\frac{1}{2},\ h_{\pi^{-}\Sigma^{+}}=1 , \
h_{\pi^{+}\Sigma^{-}}=0,  \notag \\
h_{K^-p}&=h_{\bar{K}^0n}=0  \notag \\
h_{\eta\Lambda}&=\frac{1}{3\sqrt{2}} , \
h_{\eta\Sigma^{0}}=-\frac{1}{\sqrt{6}},\
h_{K^{0}\Xi^{0}}=1,\
h_{K^{+}\Xi^{-}}=0 . \notag
\end{align}
The first term in Eq.~\eqref{eq:amplitude} stands for the direct process to generate the final state $j$, while the second term represents the rescattering effect through the final state interaction. Even for the channel which has no direct transition ($h_{j}=0$), $\mathscr{M}_{j}$ can be nonzero because of the rescattering term. The partial decay width to channel $j$ is given by integrating over the three-body phase space $d\Pi_{3}$ as
\begin{align}
\Gamma_j &=\int d\Pi_3 |\mathscr{M}_j|^2.  \label{eq:width}
\end{align}
The invariant mass distribution is calculated by differentiating both sides by $M_{\rm inv}$, leading to
\begin{align}
\frac{d\Gamma_j}{dM_{\rm inv}} &=\frac{1}{(2\pi)^3}\frac{p_{D^{0}}\tilde{p}_jM_{\Xi_b^0}M_j}{M_{\Xi_b^0}^2} |\mathscr{M}_j|^2,  \label{eq:massdis}
\end{align}
with
\begin{align}
p_{D^{0}}=& \frac{\lambda^{1/2}(M_{\Xi_b^0}^2,m_{D^{0}}^2,M_{\rm inv}^2)}{2M_{\Xi_b^0}},\ \tilde{p}_j= \frac{\lambda^{1/2}(M_{\rm inv}^2,M_j^2,m_j^2)}{2M_{\rm inv}},  \notag \\
&\lambda(x,y,z) = x^{2}+y^{2}+z^{2}-2xy-2yz-2zx.  \notag
\end{align}

\section{Numerical results}   \label{results}  

\begin{figure}[tb]
\begin{center}
\includegraphics[width=8cm,bb=0 0 846 594]{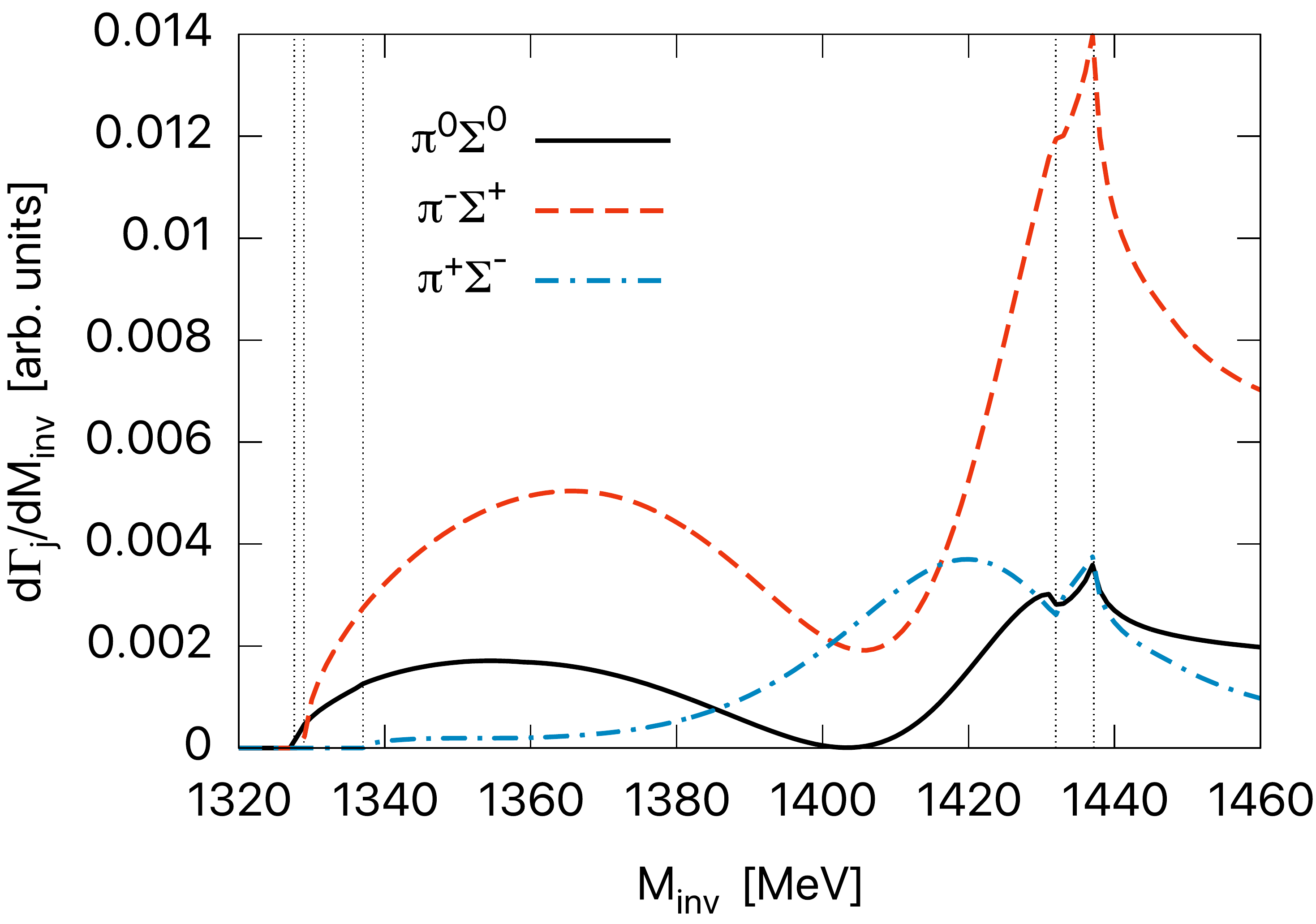}
\caption{Invariant mass distributions of the $\Xi_{b}\to D^{0}(MB)_{j}$ decay. Solid, dotted, and dash-dotted curves represent the $\pi^{0}\Sigma^{0}$, $\pi^{-}\Sigma^{+}$, and $\pi^{+}\Sigma^{-}$ final states, respectively. The vertical dotted lines stand for the threshold energies of the $\pi\Sigma$ and $\bar{K}N$ channels.}
\label{fig:massdis}  
\end{center}
\end{figure}

We calculate the invariant mass distribution of the $\Xi_{b}\to D^{0}(MB)_{j}$ decay for the $\pi\Sigma$ final states in the $\Lambda(1405)$ energy region, expecting that the spectrum reflects the $\pi\Sigma\to\pi\Sigma$ amplitude. Because the factor $V_{P}$ only scales the magnitude, we show the line shapes of the $\pi\Sigma$ spectra in arbitrary units in Fig.~\ref{fig:massdis}. While the $\pi^{+}\Sigma^{-}$ spectrum shows a peak structure in the $\Lambda(1405)$ region, the peaks in the $\pi^{0}\Sigma^{0}$ and the $\pi^{-}\Sigma^{+}$ channels appear in much lower energy region. Instead, these spectra have a dip in the $\Lambda(1405)$ region, and the threshold cusp at the $\bar{K}N$ threshold is enhanced in the $\pi^{-}\Sigma^{+}$ spectrum.

%
%

In general, the peak positions of the $\pi\Sigma$ line shapes can be shifted, due to the double-pole nature of the $\Lambda(1405)$~\cite{Jido:2003cb}. However, the pole positions of the meson-baryon amplitude $T_{ij}$ are $z_{1}=1424 -26i$ MeV and $z_{2}=1381-81i$ MeV~\cite{Ikeda:2011pi,Ikeda:2012au}, and the peak positions in Fig.~\ref{fig:massdis} are clearly lower than the pole positions. It is also known that the charged $\pi\Sigma$ spectra suffer from the interference effect between the $I=0$ and $I=1$ contributions~\cite{Nacher:1998mi}, which has been experimentally confirmed in the photoproduction $\gamma p\to K^{+}\Lambda(1405)$ by the CLAS Collaboration~\cite{Moriya:2013eb}. However, the shift of the peak position in Fig.~\ref{fig:massdis} is again much larger than what is expected from the interference with the $I=1$ component. Moreover, the shift of the peak in the $\pi^{0}\Sigma^{0}$ channel, which has no $I=1$ component, cannot be explained by the interference with the $I=1$ amplitude.

\begin{figure}[tb]
\begin{center}
\includegraphics[width=8cm,bb=0 0 846 594]{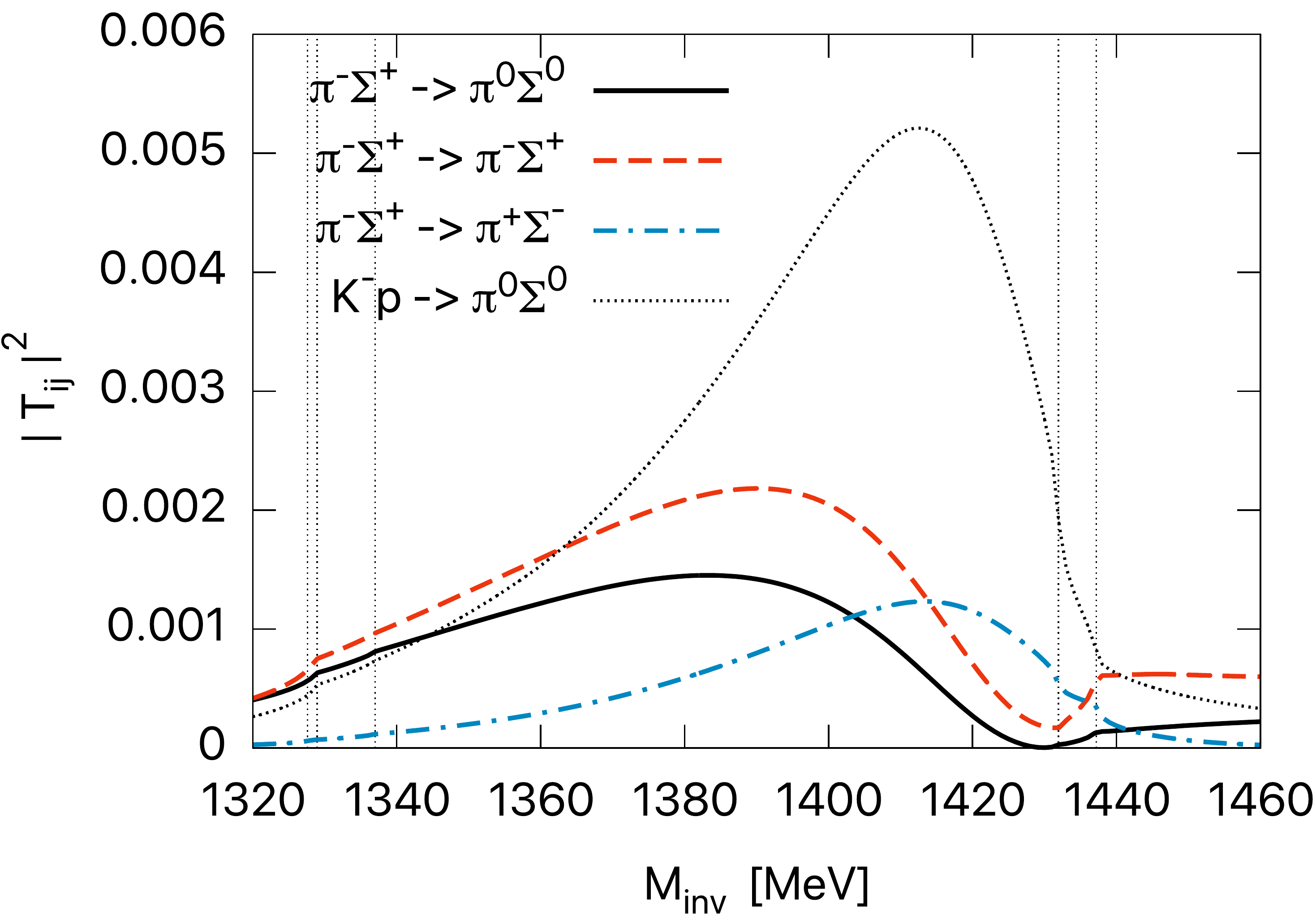}
\caption{$\pi\Sigma$ spectra from the two-body scattering amplitudes $|T_{ij}|^{2}$ in Refs.~\cite{Ikeda:2011pi,Ikeda:2012au}. Solid, dotted, and dash-dotted curves represent the $\pi^{-}\Sigma^{+}\to\pi^{0}\Sigma^{0}$, $\pi^{-}\Sigma^{+}\to\pi^{-}\Sigma^{+}$, and $\pi^{-}\Sigma^{+}\to\pi^{+}\Sigma^{-}$ processes, respectively. 
For comparison, we show $|T_{\pi^0\Sigma^0,K^-p}|^2$ by the dotted curve.
The vertical dotted lines stand for the threshold energies of the $\pi\Sigma$ and $\bar{K}N$ channels.
}
\label{fig:amplitude}  
\end{center}
\end{figure}

The above discussion is based on the interference effect for the two-body meson-baryon amplitude $T_{ij}$. In the present case, there exists additional interference effect in Eq.~\eqref{eq:amplitude}, the interference between the direct term ($h_{j}$) and the rescattering term ($h_{i}G_{i}T_{ij}$). This can be the origin of the shift of the peak positions in Fig.~\ref{fig:massdis}. To confirm this fact, we show the $\pi\Sigma$ spectra from the meson-baryon scattering amplitude $|T_{ij}|^{2}\sim\text{Im~}T_{ij}$ in Fig.~\ref{fig:amplitude}. In this figure, the peak positions are shifted from each other, but within the energy region of 1380--1420 MeV where the poles of the $\Lambda(1405)$ are located. This indicates that the interference effect with the direct process changes the peak into the dip around the 1400 MeV in Fig.~\ref{fig:massdis}, and hence the peaks of the $\pi^{0}\Sigma^{0}$ and $\pi^{-}\Sigma^{+}$ spectra appear to be shifted to the lower energy region. The ``peaks'' in Fig.~\ref{fig:massdis} therefore do not directly represent the $\Lambda(1405)$ property. The $\pi^{+}\Sigma^{-}$ spectrum is not affected by this mechanism due to the absence of the direct process ($h_{\pi^{+}\Sigma^{-}}=0$). In other words, the observation of the $\Lambda(1405)$ peak only in the $\pi^{+}\Sigma^{-}$ spectra would support the dominance of the quark-line diagram of the weak decay process we consider.

%
%

It is worth comparing the present result with the $\Lambda_{c}\to \pi^{+}MB$ reaction studied in Ref.~\cite{Miyahara:2015cja}. There, the weight $h_{i}$ for the $\pi\Sigma$ vanishes and the final state interaction is dominated by the $\bar{K}N\to\pi\Sigma$ amplitude. While a similar interference effect of the direct and rescattering processes should occur, the resulting $\pi\Sigma$ spectra properly reflects the $\Lambda(1405)$ property in the two-body amplitude. This is because the $\bar{K}N\to\pi\Sigma$ amplitude is much stronger than the $\pi\Sigma\to\pi\Sigma$ amplitude (see dotted curve in Fig.~\ref{fig:amplitude}), and the interference effect is less prominent in the $\Lambda_{c}\to \pi^{+}MB$ process. In other words, the interference effect is much enhanced in the present case, because of the small magnitude of the $\pi\Sigma\to\pi\Sigma$ amplitude. We also note that the interference of the direct and rescattering terms can in principle occur in any reactions to produce the $\Lambda(1405)$, because the two-body scattering experiment is not possible. This effect may be related to the downward shift of the peak position of the $\Lambda(1405)$ observed in the $pp\to K^{+}p\pi\Sigma$ reaction by the HADES Collaboration~\cite{Agakishiev:2012xk}.
Work in this direction, showing the dominance of the initial $\pi\Sigma$ channel, has been done in Ref.~\cite{Bayar:2017svj}.

\section{Summary}

In this paper, we have studied the decay of $\Xi_b^{0}$ into the $D^{0}MB$ state, focusing on the mechanism of the weak decay and the final state interaction between the meson-baryon pair. We show that, for the kinematics of the $MB$ pair in the $\Lambda(1405)$ energy region, the dominant contribution to the weak decay generates the superposition of the meson-baryon states in which the $\bar{K}N$ channel is absent. Hence, this process is useful to investigate the $\pi\Sigma$ originated $\Lambda(1405)$ formation amplitude. We note that this conclusion follows from the theoretical argument based on the quark-line diagram and the hadronization from a flavor singlet $\bar{q}q$ pair, and therefore experimental verification of this mechanism should be carried out in the future.

We calculate the invariant mass distribution of this decay, using the realistic meson-baryon scattering amplitude in the chiral unitary approach for the final state interaction. We find that the $\pi\Sigma$ spectra can be distorted by the interference effect between the direct and rescattering processes, except for the $\pi^{+}\Sigma^{-}$ channel. The detailed analysis of the mass distribution would enrich our understanding of the weak decay mechanism.

Our study indicates that the peak of the $\pi\Sigma$ invariant mass distribution does not always reflect the $\Lambda(1405)$ property, when the interference effect between the direct and rescattering processes is important. This implies that a naive fit to the peak of the $\pi\Sigma$ spectra by $d\Gamma_j/dM_{\rm inv}\sim |T_{ij}|^{2}$ is not always valid to extract the information of the $\Lambda(1405)$. Instead, detailed analysis of the reaction mechanism~\cite{Roca:2013av,Roca:2013cca,Mai:2014xna} is necessary to extract the information of the $\Lambda(1405)$. In this respect, the weak decay of heavy hadrons provides a new opportunity to study the property of the $\Lambda(1405)$.

\section{Acknowledgements} 
The authors are grateful to Eulogio Oset for fruitful discussions and useful comments on the manuscript. This work is partly supported by JSPS KAKENHI Grants No. 17J11386 and No. 16K17694 and by the Yukawa International Program for Quark-Hadron Sciences (YIPQS).




\end{document}